\documentclass{iaus}
\usepackage{graphicx}

\title[The X-ray Evolution of Merging Galaxies]  
{The X-ray Evolution of Merging Galaxies}
\author[Brassington, Read, Ponman]  
{Nicola J. Brassington,$^1$ Andrew M. Read$^2$ \break and Trevor J. Ponman$^1$}

\affiliation{$^1$School of Physics and Astronomy, University of Birmingham, Edgbaston, Birmingham B15 2TT, UK \break 
email:njb@star.sr.bham.ac.uk\\[\affilskip]
$^2$Department of Physics and Astronomy, University of Leicester, University Road, Leicester LE1 7RH, UK }

\pubyear{2005}
\volume{230}
\pagerange{119--126}
\date{?? and in revised form ??}
\jname{Populations of High Energy Sources in Galaxies}
\editors{Evert J.A. Meurs \& Giuseppina Fabbiano, eds.}
\begin{document}
\ifx\href\undefined\else\hypersetup{linktocpage=true}\fi 

\maketitle

\begin{abstract}

From a {\em Chandra} survey of nine interacting galaxy systems the evolution of X-ray emission during the merger process has been investigated. From comparing {\ensuremath{L_{\mathrm{X}}}}/{\ensuremath{L_{\mathrm{K}}}} and {\ensuremath{L_{\mathrm{FIR}}}}/{\ensuremath{L_{\mathrm{B}}}} it is found that the X-ray luminosity peaks $\sim$300 Myr before nuclear coalescence, even though we know that rapid and increasing star formation is still taking place at this time. It is likely that this drop in X-ray luminosity is a consequence of outflows breaking out of the galactic discs of these systems. At a time $\sim$1 Gyr after coalescence, the merger-remnants in our sample are X-ray dim when compared to typical X-ray luminosities of mature elliptical galaxies. However, we do see evidence that these systems will start to resemble typical elliptical galaxies at a greater dynamical age, given the properties of the 3 Gyr system within our sample, indicating that halo regeneration will take place within low {\ensuremath{L_{\mathrm{X}}}} merger-remnants.

\end{abstract}

\firstsection 
              
\section{Introduction}

One of the most important factors in galaxy
evolution is the effect of galaxy collisions and mergers. It is likely that most galaxies have been shaped or created through an
interaction or merger with another galaxy. In particular, many astronomers believe that the formation of many elliptical galaxies is the
product of the merger of two spiral galaxies. This was first proposed
by Toomre in 1977, illustrated by the `Toomre' sequence
(\cite{Toomre77}). This shows 11 examples from the New General
Catalogue (NGC) of Nebulae and Clusters of Stars, from approaching
disc systems to near-elliptical remnants.  The systems within this sequence have been well studied over a range of wavelengths, including X-ray.

X-ray observations are of particular importance when observing interacting galaxies as they are able to probe the dusty nucleus of the system, enabling the nature of the point source population to be established. Imaging of the the soft X-ray emission allows us to map out the diffuse gas and observe galactic-winds outflowing from the system, enabling constraints to be placed on the energetics of the outflows. From observing a selection of these interacting systems, at different stages of evolution, the processes involved in the merging of galaxies pairs can be characterised. Here we briefly describe the behaviour of the evolutionary sequence for our selected sample and discuss the mechanisms involved at key stages in this merger process.

\section{The Sample }

Our sample contains nine merging systems, selected using the following criteria;
%interacting and post-merger galaxies, selected using the following criteria;

\begin{enumerate}
\item 
All systems have been observed with {\em Chandra}.
\item 
Systems comprise of, or, originate from, two similar mass spiral galaxies.
\item
Multi-wavelength information is available for each system.
\item
The absorbing column is low, maximising the sensitivity to soft X-ray emission.
\item
A wide chronological sequence is covered; from detached pairs to merger-remnants.
\end{enumerate}

The main problem when working with a chronological study is assigning each system a merger age. This problem has been addressed through a combination of; N-body simulations, dynamical age estimates and stellar population synthesis models. From this we have established an evolutionary sequence for the nine systems in our sample, this can be seen in Table \ref{tab:gals} where; column (1) gives the system name, column (2) the distance (assuming {\ensuremath{H_{\mathrm{0}}}} = 75 km s$^{-1}$ Mpc$^{-1}$), column (3) the merger age of the system (assigning 0 Myr at the time of nuclear coalescence), column (4) the total (0.3$-$6.0 keV) X-ray luminosity ({\ensuremath{L_{\mathrm{X}}}}) of the system, column (5) the percentage of luminosity arising from the diffuse gas and column (6) the paper reference of a detailed {\em Chandra} study of the system.

\begin{table}\def~{\hphantom{0}}
  \begin{center}
  \caption{Key properties of the nine interacting systems.
  \label{tab:gals}}
  \begin{tabular}{c@{}ccc@{}ccccc}\hline

Galaxy 	& Distance	& Merger Age	& {\ensuremath{L_{\mathrm{X}}}}	& \%L$_{diff}$	& Paper reference \\
System   & (Mpc)	& (Gyr)		& ($\times10^{40}$erg s$^{-1}$) &       &                 \\\hline

Arp 270		& 28		& -0.65 		& 2.95	&	28	& \cite{Brassington05a}                           \\
The Mice	& 88		& -0.5		& 7.43	&	25	& \cite{Read03}                           \\
The Antennae	& 19		& -0.4 		& 8.40	&	45	& \cite{Fabbiano04}                           \\
Mkn 266		& 115		& -0.3		& 73.18	& 	51	& \cite{Brassington05b}                           \\
NGC 3256	& 56		& -0.2		& 90.00	&	59	& \cite{Lira02}                           \\
Arp 220		& 76		& 0 		& 26.70	&	49	& \cite{McDowell03}                           \\
NGC 7252	& 63		& 1.0 		& 6.91	&	30	& \cite{Nolan04}                           \\ 
Arp 222		& 23		& 1.2 		& 1.46	&	45	& \cite{Brassington05b}                           \\
NGC 1700	& 54		& 3.0 		& 17.58	&	83	& \cite{Diehl05}                           \\\hline
   
  \end{tabular}
  \end{center}
\end{table}

\section{The Evolution Plot}

We have obtained the B-band, K-band and FIR luminosities for each system, in addition to their {\ensuremath{L_{\mathrm{X}}}} value, as shown in Table \ref{tab:gals}. With this information we have compared the activity levels, indicated by {\ensuremath{L_{\mathrm{FIR}}}}/{\ensuremath{L_{\mathrm{B}}}} (a proxy for star-formation normalised by the galaxy mass) and {\ensuremath{L_{\mathrm{X}}}}, scaled by galaxy mass for each system. For the normalisation of {\ensuremath{L_{\mathrm{X}}}} we have used both the B-band and the K-band luminosities, {\ensuremath{L_{\mathrm{X}}}}/{\ensuremath{L_{\mathrm{B}}}} and {\ensuremath{L_{\mathrm{X}}}}/{\ensuremath{L_{\mathrm{K}}}}. This is to observe how {\ensuremath{L_{\mathrm{B}}}} varies from {\ensuremath{L_{\mathrm{K}}}}, as it is known that {\ensuremath{L_{\mathrm{B}}}} is affected by star-formation.  

\begin{figure}
 \begin{center}
 \includegraphics[angle=90,height=0.6\linewidth]{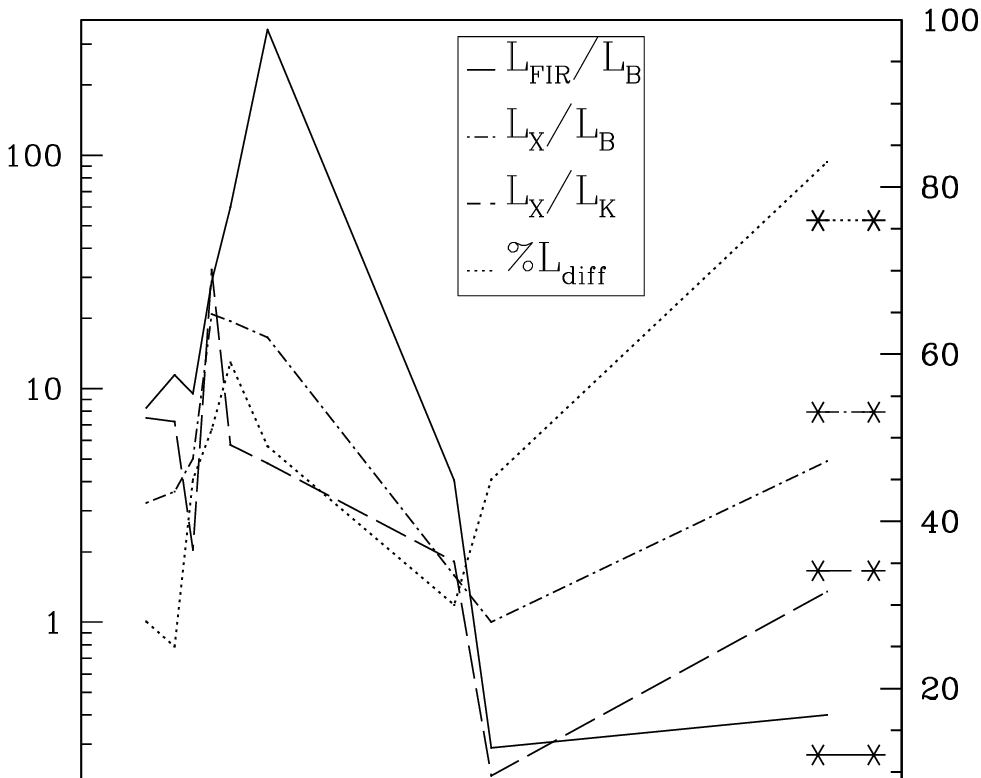} 
\caption{The evolution of X-ray luminosity in merging galaxies. Shown are {\ensuremath{L_{\mathrm{FIR}}}}/{\ensuremath{L_{\mathrm{B}}}} (solid line), {\ensuremath{L_{\mathrm{X}}}}/{\ensuremath{L_{\mathrm{B}}}} (dot-dash line), {\ensuremath{L_{\mathrm{X}}}}/{\ensuremath{L_{\mathrm{K}}}} (dashed line) and \%{\ensuremath{L_{\mathrm{diff}}}}, plotted as a function of merger age, where 0 age is defined to be the time of nuclear coalescence. All luminosity ratios ({\ensuremath{L_{\mathrm{FIR}}}}/{\ensuremath{L_{\mathrm{B}}}}, {\ensuremath{L_{\mathrm{X}}}}/{\ensuremath{L_{\mathrm{B}}}} and {\ensuremath{L_{\mathrm{X}}}}/{\ensuremath{L_{\mathrm{K}}}}) are normalised by the spiral galaxy NGC 2403. \%{\ensuremath{L_{\mathrm{diff}}}} is plotted on a linear scale, shown on the right y-axis of the image and is an absolute value. \%{\ensuremath{L_{\mathrm{diff}}}} for NGC 2403 is 12\%. The horizontal lines to the right of the plot indicate {\ensuremath{L_{\mathrm{FIR}}}}/{\ensuremath{L_{\mathrm{B}}}}, {\ensuremath{L_{\mathrm{X}}}}/{\ensuremath{L_{\mathrm{B}}}}, {\ensuremath{L_{\mathrm{X}}}}/{\ensuremath{L_{\mathrm{K}}}} and \%{\ensuremath{L_{\mathrm{diff}}}} for NGC 2434, a typical elliptical galaxy. }
  \label{fig:lbage}
 \end{center}
\end{figure}

These ratios are shown in Figure \ref{fig:lbage}, where we have plotted, not only the luminosity ratios, but also the percentage of luminosity arising from diffuse gas (\%{\ensuremath{L_{\mathrm{diff}}}}), as a function of merger age. {\ensuremath{L_{\mathrm{FIR}}}}/{\ensuremath{L_{\mathrm{B}}}} (solid line), {\ensuremath{L_{\mathrm{X}}}}/{\ensuremath{L_{\mathrm{B}}}} (dot-dash line) and {\ensuremath{L_{\mathrm{X}}}}/{\ensuremath{L_{\mathrm{K}}}} (dashed line) have been normalised by the typical spiral galaxy, NGC 2404. Whilst \%{\ensuremath{L_{\mathrm{diff}}}} (dotted line) is plotted on the right y-axis of the image, as an absolute value, where \%{\ensuremath{L_{\mathrm{diff}}}} for NGC 2403 is 12\%. The horizontal lines to the right of the plot indicate {\ensuremath{L_{\mathrm{FIR}}}}/{\ensuremath{L_{\mathrm{B}}}}, {\ensuremath{L_{\mathrm{X}}}}/{\ensuremath{L_{\mathrm{B}}}}, {\ensuremath{L_{\mathrm{X}}}}/{\ensuremath{L_{\mathrm{K}}}} and \%{\ensuremath{L_{\mathrm{diff}}}} for NGC 2434, a typical elliptical galaxy (\cite{Diehl05}). 

From this figure, it can be seen that the activity increases with merger age up to the point of nuclear coalescence. At this time the activity peaks, showing that this point of evolution is the most violent stage in the merger process. After coalescence, the activity decreases for $\sim$1.2 Gyr and then flattens. At this point it exhibits a value of activity similar to that observed in typical elliptical galaxies, as indicated by NGC 2434. 

Both of the normalised X-ray luminosity ratios, {\ensuremath{L_{\mathrm{X}}}}/{\ensuremath{L_{\mathrm{K}}}} and {\ensuremath{L_{\mathrm{X}}}}/{\ensuremath{L_{\mathrm{B}}}}, exhibit broadly the same trends, with values increasing until $\sim$300 Myr before nuclear coalescence takes place, and then dropping until $\sim$1.2 Gyr after the nuclei have coalesced.  Both of these values then begin to rise again, and, from the indication of the trend lines, seem likely to resemble an elliptical galaxy, at a later time, although this exceeds the evolution time within our sample.  \%{\ensuremath{L_{\mathrm{diff}}}} exhibits a similar trend, 
%to that of the normalised X-ray luminosity values
with the value peaking at a time before coalescence (although the peak is at $\sim$200 Myr) and then dropping, before rising once more, exhibiting a similar value to that of NGC 2434.

The results from this study are initially surprising, given that a similar study carried out with {\em ROSAT} observations (\cite{Read98}), albeit a study that does not include all the same systems as ours, found that the peak in {\ensuremath{L_{\mathrm{X}}}} was coincident with the peak in {\ensuremath{L_{\mathrm{FIR}}}}/{\ensuremath{L_{\mathrm{B}}}} and nuclear coalescence. It is likely that this difference is due to the inclusion of the merging galaxy NGC 520 (a
system which has been described as an 'anomalous half-merger' (\cite{Read05})), in the \cite{Read98} study. We suggest that the reason that {\ensuremath{L_{\mathrm{X}}}} peaks at a time before nuclear coalescence is a consequence of large-scale diffuse outflows being driven out of the interacting systems by the starbursts taking place within these galaxies. 

Starburst-driven winds are responsible for the transport of gas and energy out of star forming galaxies. The energetics of these galactic winds were investigated in \cite{Strickland00}, via two-dimensional hydro-dynamical simulations. From modelling {\ensuremath{L_{\mathrm{X}}}} they found that, once superbubbles had broken out of the galactic disc and formed galactic winds, {\ensuremath{L_{\mathrm{X}}}} dropped rapidly ($\sim$10 Myr). It was also found that the soft X-ray emission from these winds arises from the cooler, denser, low volume filling factor gas ($\eta =0.01-$2 \%) which contains only $\sim$10\% of the energy of the wind. Because of this, {\em Chandra} can only provide lower limits on the mass and energy content of the galactic winds. Consequently, more evolved, pre-merger systems that have wide-spread galactic winds exhibit lower {\ensuremath{L_{\mathrm{X}}}} than starburst systems that have yet to experience large-scale diffuse outflows.

Another interesting result from this survey is the increase in {\ensuremath{L_{\mathrm{X}}}} in the older merger-remnants. In previous studies post-merger systems have been found to be X-ray dim when compared to elliptical galaxies. In this study we have extended the merger sequence to include a 3 Gyr system, and, in doing so, have observed that these under-luminous systems seem likely to have increased {\ensuremath{L_{\mathrm{X}}}} at a greater dynamical age. Given that these merger-remnants have been shown to be quiescent, we know that this increase in {\ensuremath{L_{\mathrm{X}}}} is not due to any starburst activity within the system. Coupling this, with the increase in \%{\ensuremath{L_{\mathrm{diff}}}}, indicates that diffuse X-ray gas is being produced, leading to the creation of X-ray haloes, as observed in mature elliptical galaxies. \cite{Osul01} investigated the relationship between {\ensuremath{L_{\mathrm{X}}}}/{\ensuremath{L_{\mathrm{B}}}} and spectroscopic age in post-merger ellipticals and found that there was a long term trend ($\sim$10 Gyr) for {\ensuremath{L_{\mathrm{X}}}} to increase with time. The mechanism by which they explain the regeneration of hot gas haloes in these galaxies is one in which an outflowing wind to hydrostatic halo phase is driven by a declining SNIa rate. They argue that a scenario in which gas, driven out during the starburst, infalls onto the existing halo is not the dominant mechanism in generating X-ray haloes as this mechanism would only take $\sim$1$-$2 Gyr and would therefore not produce the long-term trend they observe. With the current information from our sample, neither scenario can be ruled out.

\section{Conclusions}

From our sample of nine interacting galaxy systems we have compared {\ensuremath{L_{\mathrm{X}}}}/{\ensuremath{L_{\mathrm{K}}}} and {\ensuremath{L_{\mathrm{FIR}}}}/{\ensuremath{L_{\mathrm{B}}}}, and shown that there is a peak in {\ensuremath{L_{\mathrm{X}}}} $\sim$300 Myr before nuclear coalescence takes place. We suggest that the reason for this decrease in {\ensuremath{L_{\mathrm{X}}}} at a time before coalescence is due to hot gaseous outflows breaking out of the galactic discs of these systems. We have also shown that the X-ray dim merger-remnant systems seem likely to evolve into more luminous systems at a greater dynamical age, given the properties observed in the 3 Gyr system within our sample. This indicates that halo regeneration will take place within low luminosity merger-remnant systems. Greater detail of this work can be found in \cite{Brassington05b}.

\end{document}